\documentstyle[12pt]{article}
\begin{document}
%%%%%%%%%%%%%%%%
\begin{titlepage}
\rightline{\vbox{\halign{&#\hfil\cr
&UCTP-105/93\cr
&\today\cr}}}
\vspace{0.5in}
\begin{center}
{\Large\bf Gauged Yukawa Matrix Models and 2-Dimensional Lattice Theories}
\\
\medskip
\vskip0.5in

\normalsize {\bf H. Hamidian$^{\rm a}$, {\bf S. Jaimungal}$^{\rm b}$,
G.W.Semenoff$^{~\rm b}$, P. Suranyi$^{\rm a}$,\\ and L. C. R.
Wijewardhana$^{\rm a}$}
\smallskip
\medskip

{ \sl $^a$Department of Physics, University of Cincinnati\\
Cincinnati, Ohio, 45221 U.S.A.\\ $^b$Department of Physics, University
of British Columbia\\ Vancouver, British Columbia, Canada V6T 1Z1
}\smallskip
\end{center}
\vskip1.0in

\begin{abstract}
We argue that chiral symmetry breaking in three dimensional QCD can be
identified with N\'eel order in 2-dimensional quantum
antiferromagnets.  When operators which drive the chiral transition
are added to these theories, we postulate that the resulting quantum
critical behavior is in the universality class of gauged Yukawa matrix
models. As a consequence, the chiral transition is typically of first
order, although for a limited class of parameters it can be second
order with computable critical exponents.
\end{abstract}

\end{titlepage}
\baselineskip=20pt

One of the most intriguing features of quantum spin systems is their
relationship to gauge theories.  This connection was originally used
to study chiral symmetry breaking in quantum chromodynamics (QCD),
where the strong coupling limit resembles a spin system ~\cite{sb}.
More recently, the analogy has been exploited to prove that certain
gauge theories break chiral symmetry in the strong coupling
limit~\cite{seiler,ls,dlss}.  It has also been used to formulate mean
field theories for magnetic systems ~\cite{mft}.  For the most part,
these works use the formal similarity between a gauge theory and a
spin system at the lattice distance scale.  Recently it has been
suggested that the analogy is much broader in that it can account for
the quasi-particle spectrum and other infrared features of the two
systems ~\cite{bpl}.  In this letter we shall present evidence for the
latter by discussing a common feature of the phase diagrams of
2-dimensional quantum antiferromagnets and 3-dimensional QCD.  The
dependence of the chiral symmetry breaking pattern on the number of
flavors and colors of quarks in QCD is similar to that of the
antiferromagnet where the rank of the spin algebra and the size of its
representation play the same role as the number of flavors and colors,
respectively.  We shall also study the critical behavior associated
with a chiral or N\'eel phase transition.  Such a transition must be
driven by operators which are added to the QCD or antiferromagnet
hamiltonian and which have the appropriate symmetries.  We argue that
these transitions fall into a universality class which can be analyzed
using the epsilon expansion.  We show that in many cases they are
fluctuation induced first order transitions.

It has been observed that, in both 2+1-dimensional QED
\cite{appelquist} and QCD ~\cite{an},  there exists a critical
number of flavors such that if $N_F<N_F^{\rm crit.}$ the model breaks
chiral symmetry spontaneously and if $N_F>N_F^{\rm crit.}$ the theory
is in a chirally symmetric, deconfined phase.  For large $N_F$ and for
large number of colors $N_C$ the equation of the critical line is
approximately $N_F-
\frac{128}{3\pi^2}N_C=0$.  A heuristic argument for this behavior is that
when $N_F>>N_C$ internal gluon exchanges and the gluon self-coupling
are suppressed by factors of $N_C/N_F$.  Resummation of leading order
diagrams, which are chains of bubbles, produces an effective
interaction which falls off like $1/r$, rather than the tree level
$\ln\vert r\vert$. The weak coupling of order $N_C/N_F$ and mild
infrared behavior of this resummed theory result in a chirally
symmetric, de-confined phase.  When $N_F$ is small, the effective
coupling is large and can generate a condensate, which is already seen
in QED \cite{appelquist}.  In fact, in QCD, when $N_F<<N_C$ all planar
diagrams contribute to processes, making the effective interaction
string-like ~\cite{th} and the theory is in a confining and chiral
symmetry breaking phase ~\cite{cw}.  Numerical simulations
{}~\cite{kogut} of 3-dimensional QED support this scenario with
$N_F^{\rm crit.}\sim 4$.

Mass operators for basic 2-component fermions in 2+1 dimensions are
pseudoscalars and break parity explicitly \cite{jt}.  Massless 2+1
dimensional QCD with an odd number of flavors of 2-component fermions
is afflicted with the parity anomaly \cite{ns} which generates a
parity violating Chern-Simons term and also fermion mass term by
radiative corrections.  With an even number of flavors, there exists a
parity and gauge invariant regularization and QCD is the
2+1-dimensional analog of a vector-like gauge theory in 3+1
dimensions.  In particular a kind of chiral symmetry can be defined.
It is known that, in this case, parity cannot be broken spontaneously
{}~\cite{vw} and therefore to study chiral symmetry breaking it is
necessary to seek parity conserving mass operators.  Following
{}~\cite{appelquist,an,pp} we shall use $N_F$ species of 4-component
fermions.  The flavor symmetry of massless QCD in this case is
actually $SU(2N_F)$.  We will add operators to the action which reduce
the symmetry to $SU(N_F)$, for example, the gauged Nambu-Jona Lasinio (NJL)
model with four-fermion interaction,
\begin{equation}
S=\int d^3x\left( \frac{1}{4}F_{\mu\nu}^2+\bar\psi\gamma_\mu
D_\mu\psi+\frac{\lambda}{2}\left( \bar\psi T^A\psi\right)^2\right)
\end{equation}
where $T^A$ is a generator of $SU(N_F)$ in the fundamental
representation.  The 4-fermi operator, which is renormalizable in the
$1/N_C$ expansion ~\cite{hssw}, can drive the chiral phase transition
with condensate $\phi^A=<\bar\psi T^A\psi>$.  The results of
{}~\cite{appelquist,an} indicate that if $N_F<N_F^{\rm crit}$ the order
persists even when $\lambda=0$.  Gauged NJL models, when analyzed by
solving the gap equation \cite{njl}, exhibit second order behavior at a
surface in the space $(N_F,N_C,\lambda)$.  Our analysis will indicate
that for a large range of parameters fluctuations make this transition
first order.  Our results do not apply to the hypothetical case where
$N_F$ or $N_C$ are varied to drive the transition \cite{atw}.

The 2-dimensional generalized antiferromagnet has Hamiltonian
\begin{equation}
H_{\rm spin}= \kappa~\sum_{<x,y>}\sum_{A=1}^{N_F^2-1}J^A(x)J^A(y)
\label{afm}
\end{equation}
with $<x,y>$ nearest neighbor sites $x$ and $y$ on a square lattice
and the spin operators $J^A(x)$ are in an irreducible representation
of the $SU(N_F)$ Lie algebra
\begin{equation}
\left[ J^A(x), J^B(y)\right]=if^{ABC}\delta(x,y)J^C(x)
\end{equation}
When the representation at each site is a rectangular Young Tableau
with $m$ rows and $N_C$ columns, it is convenient to represent the
spin operators by the fermion bilinears
\begin{equation}
J^A(x)=\sum_{\alpha=1}^{N_C}\sum_{a,b=1}^{N_F}
\psi^{\dagger}_{a\alpha}(x) T^A_{ab}\psi_{b\alpha}(x)
\end{equation}
The fermions have the anticommutator,
\begin{equation}
\{ \psi_{a\alpha}(x),\psi^{\dagger}_{b\beta}(y)\}=
\delta_{ab}\delta_{\alpha\beta}\delta(x,y)
\label{fa}
\end{equation}
Constraints which project out the irreducible representation of the
spin algebra are
\begin{equation}
{\cal G}_{\alpha\beta}(x)\equiv
\sum_{a=1}^{N_F}
 \psi^{\dagger}_{a\alpha}(x)\psi_{a\beta}(x)-\delta_{\alpha\beta}
m~\sim~0~~\forall x
\label{gl1}
\end{equation}
${\cal G}_{\alpha\beta}(x)$ obeys the $U(N_C)$ Lie algebra, commutes
with the Hamiltonian and acts as the generator of gauge
transformations with gauge group $U(N_C)$.

The critical behavior of the antiferromagnet was examined by Read and
Sachdev ~\cite{rs} using semiclassical methods.  The only free
parameters are the integers $N_C$ and $N_F$.  $N_C>>N_F$ is the
classical limit of large representations, where the classical N\'eel
ground state is stable with the staggered spin order parameter
\begin{equation}
\mu_{ab}= (-1)^{\sum_i x_i}< \sum_{\alpha=1}^{N_C}
\psi_{a\alpha}(x)\psi_{b\alpha}(x)>
\label{op}
\end{equation}
On the other hand, the limit $N_F>>N_C$ is the quantum limit where
fluctuations are important and the system is in a spin disordered
state.  For both $N_C$ and $N_F$ large, they find a line of second
order phase transitions in the $(N_C,N_F)$ plane at $N_F={\rm
const.}\cdot N_C$ where the constant is a number of order one.

The relationship between the antiferromagnet and QCD is a very close
one.  There is an argument in ref. \cite{ls} which maps the strong
coupling limit of lattice QCD onto the antiferromagnet with
Hamiltonian (\ref{afm}).  The lattice regularization of
the QCD Hamiltonian uses staggered fermions ~\cite{kog},
\begin{equation}
H=\sum_{<x,y>}\left( \psi^{\dagger}_{a\alpha}(x)U^{\alpha\beta}_{xy}
\psi_{a\beta}(y)+ h.c.
+\frac{e^2}{2} \sum_{A=1}^{N_C^2}\left( E^A(xy)\right)^2\right)+
\frac{1}{2e^2}\sum_{\Box} {\rm tr}\left(\prod_{\Box}U
+\prod_{\Box}U^{\dagger}
\right)
\end{equation}
where the first sum is over links and the second is over plaquettes
$\Box$ of the lattice.  The gauge fields, which are unitary matrices
$U_{xy}$ and electric field operators occupy links and satisfy the
algebra
\begin{equation}
\left[ E^A(xy),E^B(zw)\right]= if^{ABC}E^C(xy)\delta(xy,zw)
\end{equation}
\begin{equation}
\left[ E^A(xy), U_{wz}\right]= i\delta(xy,zw)U_{xy}T^A ~~~.
\end{equation}
The Hamiltonian is supplemented by the Gauss' law constraint
\begin{equation}
\sum_{y\in{\cal N}(x)}E^A(xy)T^A_{\alpha_\beta}+ \sum_{a=1}^{N_F}
\psi^{\dagger}_{a\alpha}(x)\psi_{a\beta}(x)-\delta_{\alpha\beta}N_F/2\sim 0
\label{gl3}
\end{equation}
which enforces gauge invariance.  Here the first summation is over
nearest neighbors of $x$.

Staggered fermions have a relativistic continuum limit when their
density is 1/2 of the maximum that is allowed by Fermi statistics and
the kinetic Hamiltonian has phases which produce an effective $U(1)$
magnetic flux $\pi$ per plaquette ~\cite{kog,ls}. In order to obtain
these phases in the continuum limit, we have chosen the sign of the
third term in the Hamiltonian so that it is minimized by the
configuration of gauge fields with the property $<\prod_{\Box}U>=-1$.
The constraint of half-filling $m=N_F/2$ is enforced by (\ref{gl3}).
The naive continuum limit yields 2+1-dimensional QCD with gauge group
$U(N_C)$ and $N_F$ species of massless four component fermions.  The
full chiral symmetry only emerges in the continuum limit.  On the
lattice, staggered fermions have a discrete remnant of chiral symmetry
(translation by one site) which forbids explicit fermion mass terms
{}~\cite{ls,dlss}.  A fermion mass term is a staggered density operator.
For example, a latticization of $\bar\psi(x)T^A\psi(x)$ is obtained
from the staggered magnetization density in eqn. (\ref{op}) as
$\sum_{ab}T^A_{ba}\mu_{ab}$.  Thus, the antiferromagnetic order
parameter and the order parameter for chiral symmetry breaking with a
flavor-vector condensate are identical.

The argument of ~\cite{ls} can be summarized as follows: The strong
coupling limit, $e^2\rightarrow\infty$ suppresses fermion propagation.
In the leading approximation, the Hamiltonian is minimized by the
states which contain as little electric field as possible and which
are compatible with the gauge constraint~(\ref{gl3}).  When $N_F$ is
even ~\cite{fn}, it is possible to solve Gauss' law with $E^A=0$.  The
occupation number of each site is $N_F/2$ and
$<(-1)^x\psi^{\dagger}_{\alpha a}\psi_{\alpha a}>=0$.  This is a
degenerate state - any gauge invariant state with $N_F/2$ fermions has
the same energy.  Because they are required to be color singlets, this
is the same set of states as occurs in the antiferromagnet when
$m=N_F/2$, i.e. in the representation of $SU(N_F)$ whose Young tableau
has $N_C$ columns and $N_F/2$ rows.  Furthermore, to resolve the
degeneracy, one must diagonalize the matrix of perturbations.  These
are non-zero only at second order and the diagonalization problem is
equivalent to solving for the ground state of the antiferromagnet
Hamiltonian (\ref{afm}) with $\kappa= t^2/e^2$.  Finally, since the
order parameters are identical, the N\'eel ordered states of the
antiferromagnet correspond to chiral symmetry breaking states of QCD.
Thus, the infinite coupling limit of QCD is identical to the
antiferromagnet.  A main difference between QCD with finite coupling
and the antiferromagnet is that QCD contains electric and gauge fields
which allow a fermion kinetic energy and still retain gauge
invariance, whereas in the antiferromagnet, the fermions are not
allowed to move.  One could regard the corrections to the strong
coupling limit of QCD as the addition of degrees of freedom and gauge
invariant perturbations in the antiferromagnet which allow fermion
propagation.  In fact, ref.  ~\cite{bpl} suggests even a stronger
correspondence, that the additional degrees of freedom are generated
dynamically.

A common feature of QCD and the antiferromagnet is that, aside from
$N_F$ and $N_C$ they have no free parameters.  We could imagine adding
operators of the sort that, if their coupling constant is varied, it
can induce the chiral transition.  It is tempting to speculate that
these transitions fall into a universality class which can take into
account all such modifications, as long as they respect the symmetries
of the theory.  Here, we shall restrict our attention to those which
lead to a Lorentz invariant continuum limit.  We argue that the
universality class is described by the $4-\epsilon$ dimensional
Euclidean field theory,
\begin{eqnarray}
S&=&\int d^{4-\epsilon}x\Bigg( \frac{1}{2}{\rm
tr}\nabla\phi\cdot\nabla\phi + {8\pi^2\mu^{\epsilon}\over
4!}\Big(\frac{g_1}{N_F^2}({\rm tr}\phi^2)^2+\frac{g_2}{N_F}{\rm
tr}\phi^4 \Big)+{1\over4}{\rm tr}F_{\mu\nu}^2 \nonumber\\ &&\;\;
+\bar\psi\Big(\gamma\cdot\nabla+i\mu^{\epsilon/2}e_1\gamma\cdot A
+i\mu^{\epsilon/2}e_2 \gamma\cdot TrA +
\frac{\pi\mu^{\epsilon/2}y}{\sqrt{N_FN_C}}\phi\Big)\psi
\Bigg)
\label{action}
\end{eqnarray}
The scalar  $\phi$ is an $N_F\times N_F$ traceless Hermitean
matrix.  The 4-component spinor $\psi$ is an $N_F\times N_C$ complex
matrix and $A_\mu$ is a $U(N_C)$ gauge field.  In four dimensions this
model has Euclidean Lorentz invariance, C,P and T, discrete chiral
symmetry, ($\psi\rightarrow\gamma^5\psi$, $\phi
\rightarrow -\phi$) and global $SU(N_F)$ flavor.
(\ref{action}) includes all operators which are marginal when $D=4$.

The evidence that (\ref{action}) describes the universality class
comes from previous work where we examined a similar model where gauge
couplings are absent ~\cite{hssw}. We showed that the anomalous
dimensions of operators computed in the model (\ref{action}) with
$e_i=0$ were identical to leading order in $1/N_C$ and $\epsilon$ to
those of a $2<D<4$ dimensional four-fermi theory.  That a $4-\epsilon$
dimensional Yukawa-Higgs theory has the same universal critical
behavior as lower dimensional four-fermi theories with the same
symmetries was originally suggested by Wilson ~\cite{wilson}.  For the
case $N_F=1$, where the chiral symmetry is discrete, higher order
computations have been carried out ~\cite{kllp,ryk}. The results, as
well as those of lattice simulations, support the universality
hypothesis ~\cite{kllp}.  We conjecture that (\ref{action}) represents
the universality class of lower-dimensional four-fermi theories with
$U(N_C)$ gauge invariance.  We shall show that, as a consequence, for
a large range of values of $(N_C,N_F)$, the chiral phase transition is
a fluctuation induced first order transition.  When it is second
order, critical exponents are in principle computable in the epsilon
expansion.

In order to analyze (\ref{action}), we use the $\epsilon$ expansion
and the renormalization group.  The 1-loop beta functions to first
order in $\epsilon$ are,
\begin{eqnarray}
\beta_1&=&-\epsilon g_1+\frac{N_F^2+7}{6N_F^2}g_1^2+
\frac{2N_F^2-3}{3N_F^2}g_1g_2+\frac{N_F^2+3}{2N_F^2}g_2^2
+\frac{1}{2N_F}y^2g_1
\nonumber\\
\beta_2&=& -\epsilon g_2+\frac{2}{N_F^2}g_1g_2+\frac{N_F^2-9}{3N_F^2}g_2^2
-\frac{3}{8N_CN_F}y^4+\frac{1}{2N_F}y^2g_2
\nonumber\\
\beta_y &=& -\frac{\epsilon}{2}y- \frac{3}{16\pi^2} \frac{N_C^2-1}{N_C} e_1^2 y
- \frac{3}{8\pi^2} e_2^2y
+ \frac{N_F^2 + 2N_F N_C -3 }{16N_F^2 N_C} y^3
\nonumber\\
\beta_{e_1} &=& -\frac{\epsilon}{2} e_1 - \frac{11N_C-2N_F}{48\pi^2} e_1^3
{}~~,~
\beta_{e_2} = -\frac{\epsilon}{2} e_2 +  \frac{N_CN_F}{12\pi^2}e_2^3
\label{beta}
\end{eqnarray}
We can also compute the anomalous dimensions of the scalar and the
fermion field.  When $N_C$ and $N_F$ are large we obtain
\begin{equation}
\Delta_S=1-\frac{2\gamma^2-11\gamma-18}{(2\gamma-11)(\gamma+2)}
\frac{\epsilon}{2}
\end{equation}
\begin{equation}
\Delta_F=\frac{3}{2}-\frac{2\gamma^2-15\gamma-50}{(\gamma+2)(2\gamma-11)}
\frac{\epsilon}{4}
\end{equation}
respectively, where $\gamma\equiv N_F/N_C$.  We shall see that the
region where there can be second order behavior is $\gamma<8.3$.
This result is reliable for small epsilon.

Fixed points occur at zeros of the beta functions, $\beta_i(g^*)=0$.
They are infrared (IR) stable or ultraviolet (UV) stable if all
eigenvalues of the stability matrix, $\partial\beta_i/\partial
g_j\vert_{g=g^*}$, are either positive or negative, respectively.  The
fixed points of the Higgs model $(e_1=e_2=y=0)$ were analyzed by
Pisarski \cite{p} and of the Yukawa-Higgs model $(e_1=e_2=0)$ in ref.
\cite{hssw}.

Second order phase transitions are possible when renormalization group
trajectories flow to an IR stable fixed point.  It is at
the IR stable fixed points that the conformal field theory with all
dimensional constants except the renormalization scale are set to
zero, exists.

If there are no such fixed points, the only possible phase transition
is a fluctuation induced first order one.  Yamagishi~\cite{y}
formulated a criterion for this behavior.  He showed that this occurs
when renormalization group trajectory crosses the surface in
coupling-constant space given by
\begin{equation}
{\cal P}_{i}(g,y,e)=0, i=1,2
\label{ym1}
\end{equation}
where ${\cal P}_{1}(g,y,e)=(4-\epsilon)(g_1+g_2) +\beta_1+\beta_2$ and ${\cal
P}_{2}=(4-\epsilon)(g_1 + \frac{N_F} 2 g_2) + \beta_1 + \frac{N_F} 2 \beta_2$
depending on whether $g_2>0$ or $g_2<0$   respectively. When the
renormalization group trajectory crosses these surfaces two further
conditions must be met. To ensure that the extremum is a local minimum,
rather than a maximum, it is necessary that $D_i >0$ for i=1,2 where
$D_1 = (4-\epsilon)(\beta_1+\beta_2)+\sum_i \beta_i  \partial/{\partial g_i}
{\beta_1 + \beta_2 }$ or
$D_2=(4-\epsilon)(\beta_1 + \frac{N_F}2 \beta_2) + \sum_i \beta_i
\partial/{\partial g_i} (\beta_1 + \frac{N_F}{2} \beta_2)$ depending on whether
$g_2>0$ or $g_2<0$.

In order that this minimum has lower free energy than the trivial $\phi=0$, it
is necessary that the couplings at that scale obey

\begin{equation}
g_1+g_2<0~~{\rm or}~~g_1+N_F g_2/2<0
\label{ym2}
\end{equation}

For the beta function (\ref{beta}), there are two distinct cases.
When $N_F<11N_C/2 $ the four dimensional non-Abelian gauge coupling is
asymptotically free.  The solution of $\beta_{e_1}=0$ is at $e_1=0$
and this fixed point is UV, rather than IR stable.  In this case,
there should be a nonperturbative behavior associated with confinement
which is inaccessible to our computation.  When $N_F>11N_C/2$ the four
dimensional gauge coupling is not asymptotically free.  $\beta_{e_1}$
has two zeros, the UV attractive one located at zero and the IR
attractive one at
\begin{equation}
e_1^{*2}  =  \frac{ 24\pi^2}{2N_F-11N_C} \epsilon
\end{equation}
Using this solution in $\beta_y$, we see that the Yukawa coupling
constant always has an UV attractive fixed point at zero coupling and
an IR attractive fixed point at
\begin{equation}
y^{*2}  =  \left ( 1 - 9 \frac{N_C-\frac{11}{2N_F}}{(11 N_C - 2N_F)}
\right )
\frac{8N_F^2N_C}{N_F^2+2N_F N_C-3} \epsilon
\label{yuk}
\end{equation}
Finally, using (\ref{yuk}), the equations for fixed points of the
matrix self-couplings are
\begin{eqnarray}
0&=&\left( \frac{y^{*2}}{2N_F}-\epsilon\right)
g^*_1+\frac{N_F^2+7}{6N_F^2}g^{*2}_1+
\frac{2N_F^2-3}{3N_F^2}g^*_1g^*_2+\frac{N_F^2+3}{2N_F^2}g^{*2}_2
\nonumber\\
0&=& \left( \frac{y^{*2}}{2N_F}-\epsilon\right)
g^*_2+\frac{2}{N_F^2}g^*_1g^*_2+\frac{N_F^2-9}{3N_F^2}g^{*2}_2
-\frac{3}{8N_CN_F}y^{*4}
\label{nb}
\end{eqnarray}
Numerical investigations show that IR stable fixed points (in fact
only one) exist when $(11/2)N_C<N_F<N_F^*(N_C)$.  The upper critical
$N_F^*(N_C)$ intersects the line $(11/2)N_C$ at $N_C=1$ and when $N_C$
and $N_F$ are large, $N_F^*(N_C) \approx 10.7 N_C$.

Naively one would expect second order behavior for $N_C$ and $N_F$ in
this entire region.  However, in some cases, renormalization group
trajectories can satisfy the conditions for first order behavior
before they reach the IR fixed point.  In particular, if the IR fixed
point is such that, $g_1^*+g_2^*<0$, then the far IR behaviour of the
effective potential is approximately $V[\phi]\approx(g_1^*+g_2^*)
\phi^{2(4-\epsilon)/(2-\epsilon)}$ and the trivial configuration
$\phi=0$ is a local maximum of the potential.  In addition, if this
$(g_1^*,g_2^*)$ is the limit of a flow originating from the
ultraviolet stability wedge, $\{g_1+g_2>0~ \cap ~g_1+N_Fg_2/2>0\}$,
the potential is bounded from below, and since it has a local maximum
at the origin, it must have a minimum when $\phi\neq0$. Hence, in this
case, there is always a first order phase transition, even though an
IR stable fixed point exists.

In fact, it is easy to see that the flow in this case intersects the
surface (\ref{ym1}).  Since $g_1^*+g_2^*<0$, any trajectory which
flows from the UV fixed point at the origin (in a direction within the
stability wedge) to the IR fixed point must intersect the surface
${\cal P}=0$ in (\ref{ym1}). Consider a trajectory which begins near
the UV fixed point at the origin, where the couplings are small and
$\beta_i\approx-\epsilon g_i$. Since the trajectory must be in the
region $g_1+g_2>0$, ${\cal P}\approx(4-2\epsilon)(g_1 +g_2)$ which is
positive.  At the IR fixed point, the beta functions in ${\cal P}$
vanish and it takes the form ${\cal P}=(4-\epsilon)(g_1^*+g_2^*)$
which is negative.  Since ${\cal P}(g,y,e)$ changes sign as we follow
the trajectory from the UV to the IR fixed point, it must go through
zero at least once.

We have verified numerically for a
$(N_C,N_F)=\{(2,15),(2,18),(100,1000)\}$, with $\epsilon=0.1$ that
they all satisfy the criterea for a first order phase transition.
Generally, IR fixed points with $g_1^*+g_2^*<0$ occur in the range
$N_F^{\cdot}(N_C)<N_F<N_F^{*}(N_C)$.  In this region, the theory has
an IR stable fixed point, but undergoes a first order phase
transition. For large $N_F$ and $N_C$, $N_F^\cdot(N_C)= 8.3 N_C$.

The other possibility is that the IR stable fixed point has the
property $g^*_1+g^*_2>0$. In this case, we have found that there are
two kinds of trajectories, those which flow from the UV to IR fixed
points without encountering the surface ${\cal P}=0$ and lead to
second order transitions, and those which visit the region where
$g_1+g_2<0$.  The latter trajectories cross the surface twice and
satisfy the conditions for first order behavior at one of the
intersections.  In fig.1 we have plotted several RG flows for the case
$N_C=2$, $N_F=13$ and $\epsilon=.1$. (The features of the flow are
insensitive to $\epsilon$ in the domain $.1<\epsilon<1$. Near
$\epsilon=1$, the relevant coupling constants are large and one would
not expect the 1-loop approximation that we have used to be accurate
there.) The stars indicate that the flow has crossed the surface
${\cal P}=0$ in the region defined by $D_i>0$ and (\ref{ym2}).  Hence
the transition associated with those trajectories is first order.  The
crosses denote points where the flow intersects the surface
(\ref{ym1}), but not when $g_1+g_2<0$.  They do not correspond to
global minima of the effective potential and therefore do not indicate
a first order transition.

The physical quantum spin $j$ antiferromagnet corresponds to $N_F=2$
and $N_C=2j$. With $N_C=1~ (j=1/2)$ the non-abelian field must be
removed, and the resulting theory has no IR fixed point, indicating a
first order transition. $N_F=2,N_C\ge2$ (i.e. $j\geq 1$) is in the
asymptotically free regime and hence these antiferromagnets cannot be
analyzed by our techniques. We speculate that confinement is
associated with a nonperturbative IR fixed point of the gauge
coupling.  In that case, it is likely that these antiferromagnets
would have a second order transition.

The work of G. S.  and of S. J. is supported by the Natural Science
and Engineering Research Council of Canada.  The work of H. H., P. S.
and of L. C. R. W. is supported in part by the United States
Department of Energy under grant no.  DE-FG02-84ER40153. L. C. R. W.
thanks A. Niemi for hospitality and support at the Institute for Theoretical
Physics at Uppsala where part of this research was performed.

\noindent
Figure caption: Fig. 1:Projection of the renormalization group trajectories in
five dimensional coupling constant space onto the $g_1-g_2$ plane for
$N_C=2$, $N_F=13$ and $\epsilon=.1$.  Stars indicate that the
trajectory encounters the surface ${\cal P}=0$ in the region defined
by eqns (\ref{ym2}) whereas the crosses denote points of intersection
where those criterea are not satisfied.

\begin{thebibliography}{10}
\bibitem{fn}If $N_F$ is odd, the coupling ground state breaks
chiral symmetry with a flavor singlet condensate
$<(-1)^x\psi^{\dagger}_{\alpha a}\psi_{\alpha a}>$
\cite{ls,dlss}.  However, even in that case, there is still the
question of whether there is also an $SU(N_F)$ breaking condensate,
$<(-1)^x\psi^{\dagger}T^A\psi>$.
\end{thebibliography}
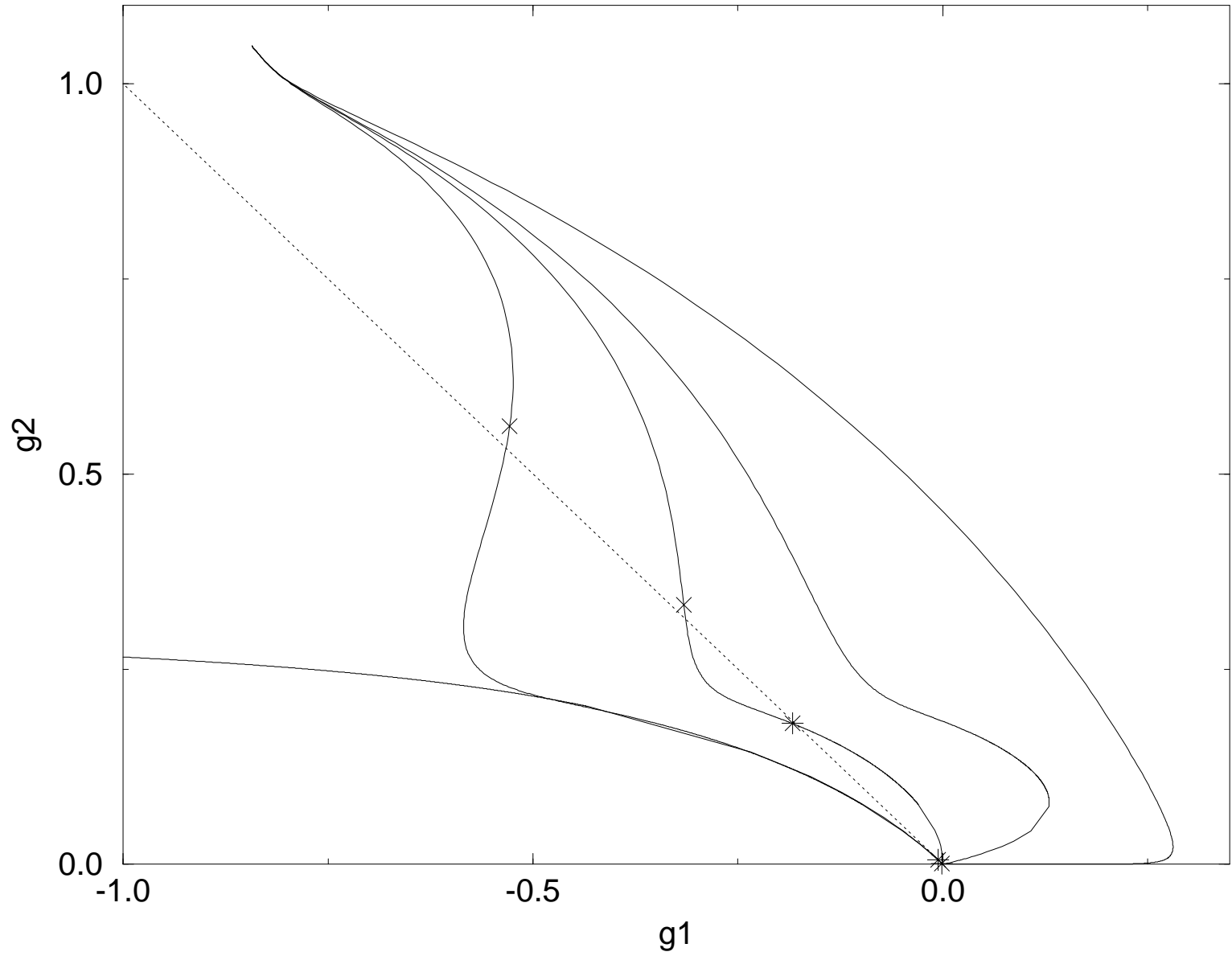
\end{document}